\documentclass[aps,prl,twocolumn,superscriptaddress,groupedaddress]{revtex4}
\usepackage[T1]{fontenc}
\usepackage[english]{babel}
\usepackage{amssymb}
\usepackage{hyperref}
\usepackage{cleveref}
\usepackage{cancel}
\usepackage{color,graphicx}
\usepackage{amsmath}
\usepackage{amsbsy}
\usepackage{amsthm}
\usepackage{bbm}
\usepackage{booktabs}
\usepackage{epsfig}
\usepackage{float}
\usepackage{graphicx}
\usepackage{dcolumn}
\usepackage{bbm}
\usepackage{color,epstopdf}
\usepackage{amscd}
\usepackage{amsfonts}  
\usepackage[]{amsmath}
\usepackage{amssymb}    
\usepackage{mathrsfs}
\usepackage{verbatim}
\usepackage[]{cases}
\usepackage{amsmath}
\usepackage{wasysym}
\usepackage[utf8]{inputenc}
\usepackage{mathtools}

\usepackage{bbold}
\usepackage{braket}
\usepackage{graphicx}
\begin{document}
\title{On the decoherence effect of a stochastic gravitational perturbation on scalar matter and the possibility of its interferometric detection}
\author{L. Asprea}
\email{lorenzo.asprea@phd.units.it}
\affiliation{Department of Physics, University of Trieste, Strada Costiera 11, 34151 Trieste, Italy}  
\affiliation{Istituto Nazionale di Fisica Nucleare, Trieste Section, Via Valerio 2, 34127 Trieste, Italy}      
\author{A. Bassi}
\affiliation{Department of Physics, University of Trieste, Strada Costiera 11, 34151 Trieste, Italy}
\affiliation{Istituto Nazionale di Fisica Nucleare, Trieste Section, Via Valerio 2, 34127 Trieste, Italy}
\author{H. Ulbricht}
\affiliation{Department of Physics and Astronomy, University of Southampton, Highfield Campus, SO17 1BJ, United Kingdom}
\author{G. Gasbarri}
\email{g.gasbarri@soton.ac.uk}
\affiliation{Department of Physics and Astronomy, University of Southampton, Highfield Campus, SO17 1BJ, United Kingdom}
\affiliation{Istituto Nazionale di Fisica Nucleare, Trieste Section, Via Valerio 2, 34127 Trieste, Italy}   
\date{\today}
\begin{abstract}
\footnotesize{We present a general master equation for the quantum dynamics of a scalar bosonic particle interacting with an external weak and stochastic gravitational field. The dynamics predicts decoherence in position as well as in momentum and energy. We show how the master equation reproduces the previous results in the literature by taking appropriate limits, thus explaining the apparent contradiction in their dynamical description. We apply our model to matter-wave interferometry, providing a practical formula for determining of the magnitude of gravitational decoherence. We compare it with the standard experimental sources of decoherence.}
\end{abstract}

\maketitle

\textit{Introduction.} - 
The recent exciting first detections of gravitational waves~\cite{ligo,neutronstar}, which marked a new era in astrophysics and cosmology, have pushed the scientific community towards the construction of ever more sophisticated ground and space based detectors~\cite{tian,decigo,et,kagra,lisa} in order to observe waves in a variety of ranges, possibly down to the cosmic background gravitational radiation. Detecting the latter would open the possiblity to gain relevant information about the universe at its very primordial stage, about $10^{-22}$ s after the Big Bang~\cite{allen}, where we expect our description of gravity to fail~\cite{qg1,qg2}, also because of its unclear relation with quantum matter.\\
In this scenario, the extreme sensitivity of matter waves~\cite{fuentes,hogan,gravwav,gao} to gravity gradients~\cite{cow} makes matter-wave interferometers a perfect candidate for exploring the gravitational wave background~\cite{allen,nonmin,astr} and, at the same time, for possibly answering some fundamental questions regarding the nature of gravity~\cite{qg,hawk,diosi,classch,nimm}, and its coupling to quantum matter.\\
In this letter we analyse the sensitivity of atom interferometry to a stochastic gravitational background, whose general effect on quantum matter is a path dependent phase shift which ultimately leads to decoherence~\cite{aha,linet}. There is a rather rich literature on the subject~\cite{sanchez,power,blencowe,breuer,ana}, which however seems to yield contradictory predictions for such an effect, in particular regarding the preferred basis of decoherence. Without a clear description of gravitational decoherence, it is not possible to asses if and to which extent matter wave interferometrs represent a viable platform to explore the gravitational background. In this letter, we present a general non relativistic model of gravitational decoherence, which clarifies the apparent discrepancies~\cite{bosons}. We show how the results in the literature can be understood as limiting case of our overarching model. In light of this general result, we characterize the study of the sensitivity of Mach-Zehnder interferometers to a stochastic gravitational background in particular regimes of interest.
 
\begin{figure}[h!]
\begin{center}
\includegraphics[scale=.117]{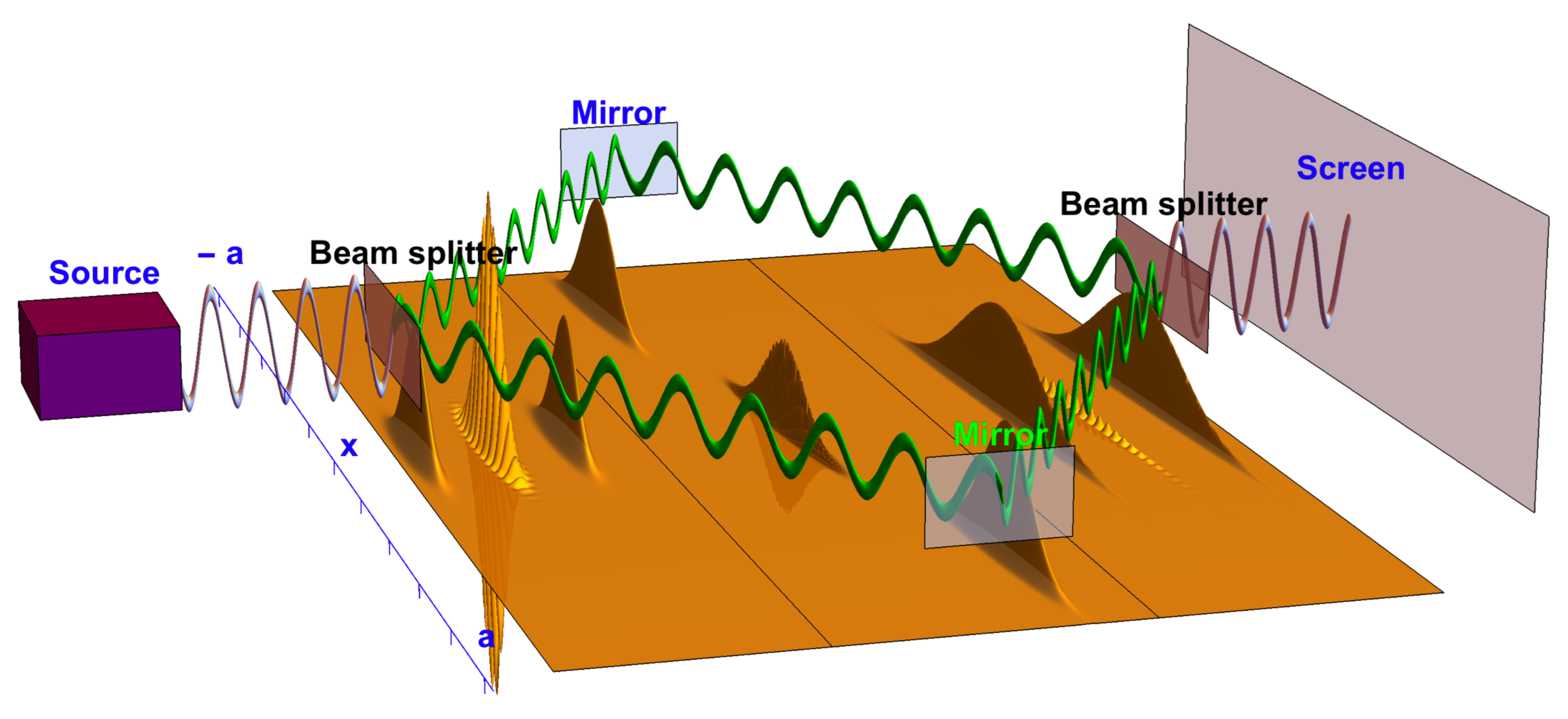}
\end{center}
\caption{Mach-Zehnder atom interferometer. A wavepacket goes through a beam splitter; the two partial waves travel the same distance before they are reflected by two mirrors to eventually collimate at a second beam splitter and be directed towards a screen where the interference fringes are observed.}
\label{f1}
\end{figure}

\begin{figure}[h!]
\centering
\textbf{Reduction in visibility from gravitational decoherence in HYPER - scalar case}\par\medskip
 \includegraphics[width=.78\linewidth]{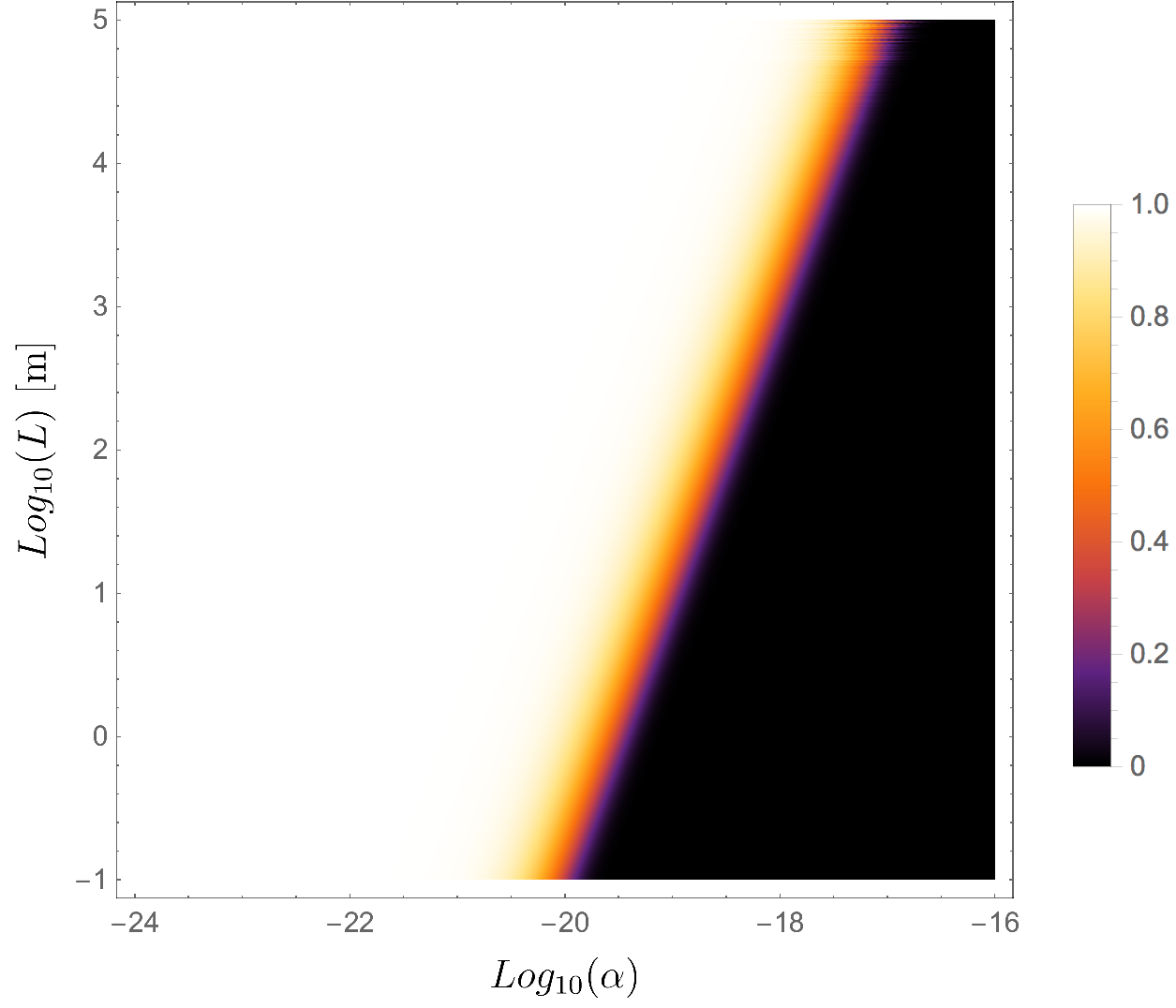}
\caption{Colored plot showing the visibility as a function of the strength $\alpha$ and of the correlation wavelength $L$ of the gravitational fluctuation.}
\label{vis}
\end{figure}

\begin{figure}[h!]
\centering
\textbf{Reduction in visibility from gravitational decoherence in HYPER - tensorial case}\par\medskip
 \includegraphics[width=.98\linewidth]{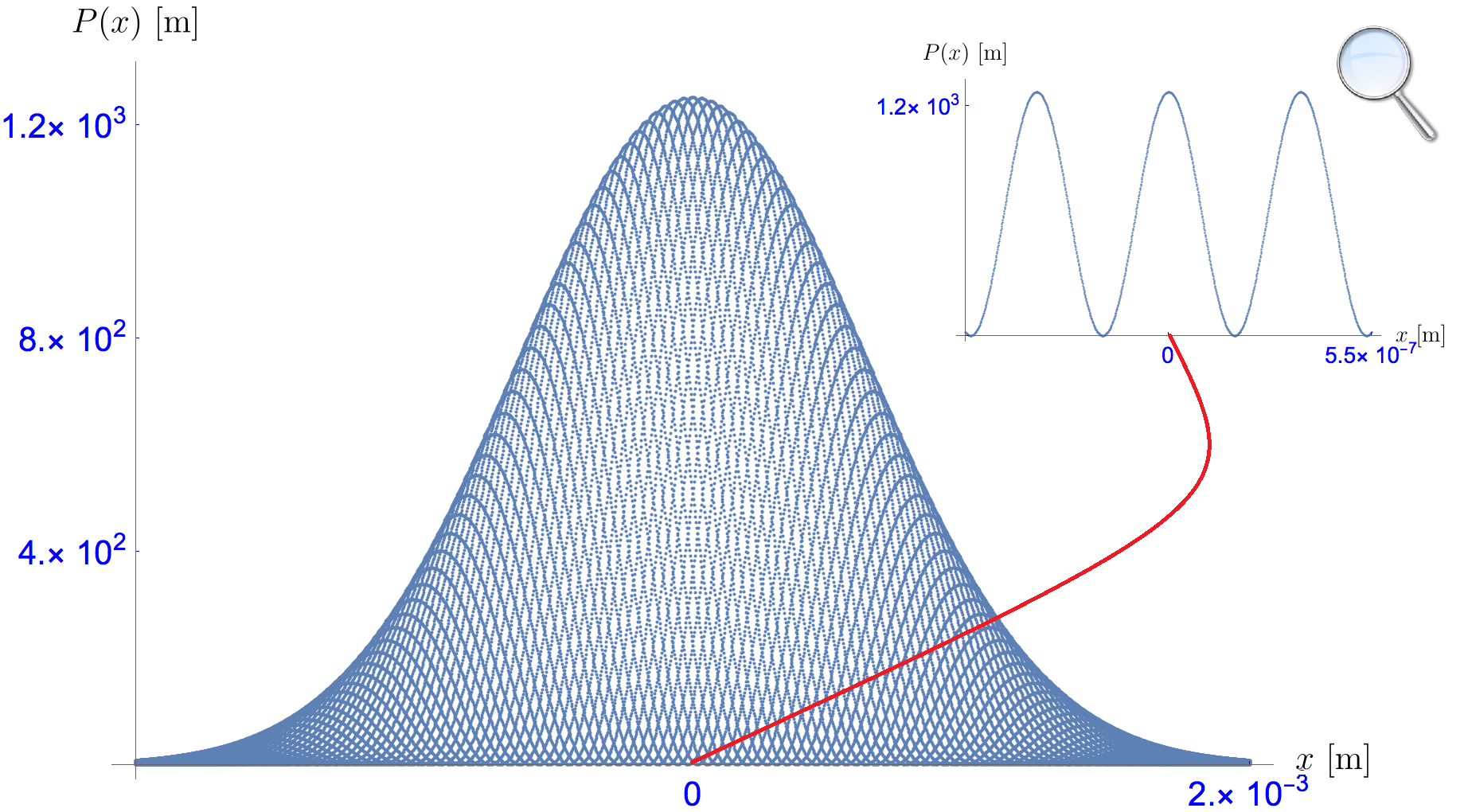}
\caption{ Probability density as a function of position ($x$). Numerical plot for all choices of $\alpha$ and $L$ in the range\\   $0 \leq \alpha \leq 10^{-5}$, $1\:\textrm{m} \leq L \leq 10^{15}\: \textrm{m}$. See \cite{supplemental} for more plots.}
\label{vism}
\end{figure}

\textit{The model.} -  In what follows we report the essential steps of the derivation of the master equation describing gravitational decohernece and its main features. We refer the reader to~\cite{bosons} for a detailed derivation which contains all calculations and references to the mathematical techniques there employed.\\
To start with, the interaction between matter and gravity is derived from the action of a scalar bosonic matter field ($\phi$) minimally coupled to the metric ($g_{\mu\nu}$). Under the assumption of small fluctuations of flat spacetime,  i.e. $g_{\mu\nu}=\eta_{\mu\nu} + h_{\mu\nu}$, $\vert h_{\mu\nu}\vert \ll 1$, the action is expanded around the Minkowski metric~\footnote{Note that by doing so, we are choosing a reference frame where macroscopic rulers and clocks are not appreciably bent by the gravitational field, as it is reasonable to happen in a real experiment.} obtaining:
\begin{align}\label{eq:lagrange}
\mathcal{S}=\int d^4x\: c^2 (\partial_{\mu}\phi^*\partial^{\mu}\phi -\frac{m^2c^2}{\hbar^2}\vert\phi\vert^2)-\frac{1}{2}h^{\mu\nu}T^{(0)}_{\mu\nu}+\mathcal{O}(h^{2})
\end{align}
from which the equations of motion (EOM) can be easily derived.
In this framework the EOM have to be understood as acting on flat spacetime, and the effect of the metric perturbation is expressed via an external force described by the coupling of the gravitational perturbation with the flat matter stress-energy tensor $(T^{(0)}_{\mu\nu})$.\\
The EOM can be rewritten in terms of the positive and negative energy components of the bosonic field, and the non relativistic limit is taken by means of the Foldy-Wouthuysen transformation~\cite{foldy}. 
The model is then extended to describe the dynamics of the center of mass of an extended body of mass $M$. The quantization follows in the canonical way.\\ 
As for the gravitational background, we specialize to the case of a gaussian stochastic perturbation around flat spacetime. After averaging over the gravitational noise, a master equation for the extended particle is derived (See Eq.(30) of \cite{bosons}).
Here we focus on the case where the stochastic perturbation is
homogeneous, isotropic and white in time , with  zero mean, and variance $\mathbb{E} [h_{\mu\nu}(\mathbf{x},t)h_{\nu\rho}(\mathbf{y},s)] = \frac{L\alpha^2}{c}u_{\mu\rho}(\mathbf{x}-\mathbf{y})\delta(t-s)$, where $\alpha$ and $L$ are respectively the strength of the fluctuations and the correlation length.
Under these assumptions, the master equation describing the dynamics of a non relativistic matter field under weak spacetime fluctuations reads:
\begin{widetext}
\begin{equation}\label{ext}
\begin{split}
\partial_t \hat{\rho} =& -\frac{i}{\hbar}\Big[\frac{\hat{\mathbf{P}}^2}{2M},\hat{\rho}(t)\Big]-\frac{\alpha^2 L c^3}{4(2\pi)^{3/2}\hbar^5 }\int d^3q\:\tilde{u}^{00}(\mathbf{q})\frac{m^2(\mathbf{q})}{M^2}\Big[\Big\{ e^{i\mathbf{q}\cdot\hat{\mathbf{X}}/ \hbar},(\frac{\hat{\mathbf{P}}^2}{4M}+\frac{Mc^2}{2})\Big\},\Big[\Big\{e^{-i\mathbf{q}\cdot\hat{\mathbf{X}}/\hbar},(\frac{\hat{\mathbf{P}}^2}{4M}+\frac{Mc^2}{2})\Big\},\hat{\rho}(t)\Big]\Big] + \\
&  -\frac{\alpha^2Lc}{4(2\pi)^{3/2}\hbar^5 }\int d^3q \:\tilde{u}^{0i}(\mathbf{q})\frac{m^2(\mathbf{q})}{M^2}\:\Big[\Big\{e^{i\mathbf{q}\cdot\hat{\mathbf{X}}/ \hbar},\hat{P}_i\Big\},\Big[\Big\{e^{-i\mathbf{q}\cdot\hat{\mathbf{X}}/\hbar},\hat{P}_i\Big\},\hat{\rho}(t)\Big]\Big]+\\
& -\frac{\alpha^2L}{4(2\pi)^{3/2}\hbar^5 c}\int d^3q \:\tilde{u}^{ij}(\mathbf{q})\frac{m^2(\mathbf{q})}{M^2}\:\Big[\Big\{e^{i\mathbf{q}\cdot\hat{\mathbf{X}}/ \hbar},\frac{\hat{P}_i\hat{P}_j}{2M}\Big\},\Big[\Big\{e^{-i\mathbf{q}\cdot\hat{\mathbf{X}}/\hbar},\frac{\hat{P}_i\hat{P}_j}{2M}\Big\},\hat{\rho}(t)\Big]\Big]
\end{split}
\end{equation}
\end{widetext}
where $\hat{\mathbf{X}}$ and $\hat{\mathbf{P}}$ are the particle's center of mass position and momentum operators, and $\tilde{u}^{\mu\nu}(\mathbf{q})$ and $m(\mathbf{q})$ are respectively the Fourier transform of the noise correlation function and of the particle mass density.
The decoherence mechanism described by Eq.~\eqref{ext} is rather complex, while the literature so far ~\cite{sanchez,breuer,blencowe,ana,power} has discussed decoherence only in momentum or position.
For this reason, in the next section we study the specific regimes in which only position or momentum decoherence dominate, and show how to recover the existing literature as limiting cases, thus reconciling apparently contradictory results.\\ 
\textit{Recovering position and momentum decoherence.} - 
The model in Eq.~\eqref{ext} describes decoherence in position when the
$h^{00}$ component of the metric fluctuations is at least of the same order of magnitude of the others, i.e.
\begin{align}\label{eq:podec}
 h^{00}\gtrsim h^{0i}, h^{ij}
 \end{align}
In this regime we are allowed to neglect the terms containing  $h^{0i}, h^{ij}$~\footnote{In the non relativistic limit $c\vert P\vert,\:\frac{P^2}{2M}\ll Mc^2$, therefore $c h^{0i}P_i,\:h^{ij}\frac{P_iP_j}{2M},\:h^{00}\frac{P^2}{2M} \ll h^{00}Mc^2 $}, thus Eq.~\eqref{ext} simplifies to:
\begin{align}\label{sgp}
&\partial_t \hat{\rho} =  -\frac{i}{\hbar}\Big[\frac{\hat{\mathbf{P}}^2}{2M},\hat{\rho}(t)\Big]\nonumber\\
& -\frac{\alpha^2 Lc^3}{(2\pi)^{3/2}\hbar^5}\int\! d^3q\, \tilde{u}^{00}(\mathbf{q})m^2(\mathbf{q})\Big[e^{i\mathbf{q}\cdot\hat{\mathbf{X}}/\hbar},\Big[e^{-i\mathbf{q}\cdot\hat{\mathbf{X}}/\hbar},\hat{\rho}(t)\Big]\Big]
\end{align}
Eq.~\eqref{sgp} describes indeed a position decoherence process; it also replicates the models in~\cite{sanchez} and in~\cite{power} under the assumptions of a pointlike particle: $m(\mathbf{r}) =M \delta^3(\mathbf{r})$ and of gaussian shaped correlation function:
\begin{align}
\label{corr}
\tilde{u}^{00}(\mathbf{q}) &= L^{3}\hbar^3e^{-\mathbf{q}^2L^2/(2\hbar^2)}
\end{align}
Choosing instead the following form for the correlation function:
\begin{equation}
u^{00}(\mathbf{x}-\mathbf{x^\prime}) = L^3\delta^3(\mathbf{x}-\mathbf{x^\prime})
\end{equation}
 one can re-obtain the model in~\cite{blencowe}  with a minor mismatch in the rate functions. 
Such a mismatch can be accounted to a different treatment of the gravitational perturbation in the two models; in~\cite{blencowe} the perturbations are described  by a quantum noise, thus allowing for complex correlation functions, while in our case the gravitational noise is classical.\\
The master equation in Eq.~\eqref{ext} describes decoherence in momentum when the correlation length of the noise is much bigger than the particle's spatial coherence.
In this regime there is a low-momentum transfer from the noise to the matter field, and we are allowed to make the following approximation $e^{i\mathbf{q}\cdot\hat{\mathbf{X}}/\hbar}\sim \hat{\mathbb{1}}$ to simplify Eq.~\eqref{ext} as follows: 
\begin{align}\label{momentum}
\partial_t \hat{\rho} = & -\frac{i}{\hbar}\Big[\frac{\hat{\mathbf{P}}^2}{2M},\hat{\rho}(t)\Big]\nonumber\\
-& \frac{\alpha^2 L}{(2\pi)^{3/2}\hbar^5 c}\int d^3q \:\tilde{u}^{00}(\mathbf{q})\frac{m^2(\mathbf{q})}{M^2}\:\Big[\frac{\hat{\mathbf{P}}^2}{2M},\Big[\frac{\hat{\mathbf{P}}^2}{2M},\hat{\rho}(t)\Big]\Big]\nonumber\\
-&  \frac{\alpha^2Lc}{(2\pi)^{3/2}\hbar^5 }\int d^3q \:\tilde{u}^{0i}(\mathbf{q})\frac{m^2(\mathbf{q})}{M^2}\:\Big[\hat{P}_i,\Big[\hat{P}_i,\hat{\rho}(t)\Big]\Big]\nonumber\\
-& \frac{\alpha^2L}{(2\pi)^{3/2}\hbar^5 c}\int d^3q \:\tilde{u}^{ij}(\mathbf{q})\frac{m^2(\mathbf{q})}{M^2}\:\Big[\frac{\hat{P}_i\hat{P}_j}{2M},\Big[\frac{\hat{P}_i\hat{P}_j}{2M},\hat{\rho}(t)\Big]\Big]
\end{align}
Eq.~\eqref{momentum} describes indeed a momentum (and energy) decoherence process.
This equation replicates the model in~\cite{breuer}, for a gaussian mass density distribution: \\
$m(\mathbf{r}) = \frac{m}{(\sqrt{2\pi}R)^3}e^{-\mathbf{r}^2/(2R^2)}$ with: $ h^{ij}\gg h^{0i}, h^{00}$ and correlation function \begin{align}\label{corrmom}
     \tilde{u}^{ij}(\mathbf{q}) = \delta^{ij}L^3\hbar^3e^{-\mathbf{q}^2L^2/(2\hbar^2)}
 \end{align}
It also recovers the result in~\cite{ana} with a minor difference in the rate function, that can again be accounted to the quantum treatment of the gravitational noise in~\cite{ana}. \\
Our general result shows that the effect of space-time fluctuations on a  non relativistic quantum matter can result both in position and/or momentum decoherence depending on the properties of the noise relative to the state of the particle. It also sets the regimes of validity of the models in the existing literature, thus solving the preferred basis puzzle.  \\
Next, we investigate the sensitivity of a Mach-Zehnder matter wave interferometer to spacetime stochastic fluctuations. We focus on the case of perturbations with large correlation length, as this limiting case embeds both the pure positional and energy decoherence effects, described respectively by Eq.(\ref{sgp}) and Eq.(\ref{momentum}) with $h^{ij}\gg h^{0i}, h^{00}$. Such effects are also of particular experimental interest as they are induced when only respectively the scalar or the tensorial component of the gravitational perturbation~\cite{weinberg} is dominant. We apply the result to a setup like HYPER~\cite{hyper}, a space based, Cesium atom interferometer aiming at testing the weak equivalence principle and measuring the Lense-Thirring effect in the Earth's gravitational field. \\
\textit{Application: Mach-Zehnder interferometry} - In a Mach-Zehnder interferometer, like the one schematically depicted in Fig.(\ref{f1}), the effect of decoherence is a loss of contrast in the interference pattern produced at the detector~\cite{gallis,joos}.
To quantify this loss we use the interferometric visibility $\nu$, which is defined in terms of the maximum ($P_\textrm{max}$) and minimum ($P_\textrm{min}$) intensity of the interference pattern: 
\begin{align}\label{eq:vis}
\nu= \frac{P_\textrm{max}-P_\textrm{min}}{P_\textrm{max}+P_\textrm{min}}
\end{align}
We therefore implement a model for the evolution of the probability density to then determine the visibility. 
Motivated by experimental interest, we consider only the pure position ($\mathcal{P}$) and energy ($\mathcal{E}$) decoherence cases in the long correlation length regime described by the simpler Eqs.(\ref{sgp}, \ref{momentum}). Furthermore, we take the spatial correlation function of the noise to be a gaussian, as in Eq.(\ref{corr}, \ref{corrmom}).\\
We restrict the analysis to point like particles, as Mach-Zehnder interferometry is currently bound to neutrons and atoms due to technical limitations~\cite{talbot}. 
 We also assume the matter-wave to be collimated and the interaction time with the mirrors to be negligible, thus we can rely on the longitudinal-eikonal approximation and reduce the study to a one dimensional problem along the transverse axis of propagation, i.e. the $x$-axis in Fig.(\ref{f1}).\\ 
We work with the characteristic function~\cite{wigner,wigrep}, which is defined in terms of the statistical operator $\rho_t$ as:
\begin{align}
\chi_t(s,q) =\frac{1}{h}\text{Tr} [e^{i(\hat{x}q-\hat{p}s)/\hbar} \hat{\rho}_{t}]
\end{align}
and is connected to the probability density, and thus the interference fringes, through the relation:
\begin{equation}
P_t(x) = \frac{1}{h}\int \chi_t(0,q) e^{-iqx/\hbar} dq
\end{equation}
In this formalism the free evolution simply reads: $\chi_t(s,q) = \chi_0(s-\frac{q}{m}t,q)$. The effect of position decoherence is described by the convolution in the $s'$ variable of the characteristic function with the kernel \cite{nimmrichter2014macroscopic}  $R_{\mathcal{P},t}(s',q)= \exp \Big[ \int_0^t ( \Gamma_{\mathcal{P}}(s'-\frac{q}{m}\tau)d\tau \Big]\delta(s')$, where $\Gamma_\mathcal{P}$ is the position decoherence rate function, which in our study is:
\begin{align}
    \Gamma_\mathcal{P}(x,x') =\frac{2 m^2\alpha^2c^3L}{\hbar^2}\left( e^{-\frac{(x-x')^2}{2L}}-1\right)
\end{align}
The effect of energy decoherence is described by the convolution of the characteristic function with the different kernel \cite{supplemental} $R_{\mathcal{E},t}(s',q)= \frac{1}{h}\int dp\: e^{\Gamma_\mathcal{E}(p-\frac{q}{2},p+\frac{q}{2})t}e^{\frac{i}{\hbar}p(s')}$,  where $\Gamma_\mathcal{E}$ is the energy decoherence rate function, which in our study is:
\begin{align}
    \Gamma_\mathcal{E}(p,p') = - \frac{\alpha^2 L (p^2-p'^2)^2}{4 M^2 \hbar^2 c}
\end{align}
Thus, the evolution of the system from the first beam splitter to the mirrors is described by:
\begin{align}\label{evo}
\chi_{t}(s,q)= \int ds' R_{i,t}(s'-s,q)\chi_{0}(s'-\frac{q}{m}t,q) 
\end{align}
where $i$ stands for $\mathcal{P}$ or $\mathcal{E}$ depending on what kind of decoherence (position or energy) occurs.\\
We model the reflection at the mirrors, following the principles of the "image charge", as the sudden transformation~\footnote{For the sake of clarity this corresponds to $\psi_{t_{\textrm{ref}}}(x)\rightarrow \psi_{t_{\textrm{ref}}}(2a-x) + \psi_{t_{\textrm{ref}}}(-2a-x)$ ; $p\rightarrow -p$ in terms of the wavefunction.}
\begin{align}
\label{eq:mirr}
\chi_{t_{\textrm{ref}}}(s,q)\rightarrow &\: \chi_{t_{\textrm{ref}}}(-4a-s,-q)+ \chi_{t_{\textrm{ref}}}(4a-s,-q) \nonumber\\
& + 2\cos\Big(\frac{a q}{\hbar}\Big) \chi_{t_{\textrm{ref}}}(-s,-q)
\end{align}
 where $2a$ is the distance between the two mirrors. Note that this applies to both the positional and the energy decoherence cases. Finally, the evolution from the mirrors to the second beam splitter is described again by means of Eq.~\eqref{evo} in the two respective cases \cite{supplemental}.\\
 In the case of the position decoherence process, this results in the following interference pattern at the screen:
\begin{align}\label{probp}
P^{(\mathcal{P})}_\textrm{scr}(x)=& \frac{1}{h} \int dq \: e^{\frac{i}{\hbar}qx} e^{\int_{0}^{t_\textrm{ref}}(\Gamma_\mathcal{P}(\frac{q}{m}\tau))d\tau}\nonumber\\
&\Big[e^{\int_{0}^{t_\textrm{ref}}\Gamma_\mathcal{P}(\frac{q}{m}\tau+4a)d\tau}\chi_{0}\big(-4a-\frac{2aq}{k},-q\big)\nonumber\\
&+e^{\int_{0}^{t_\textrm{ref}}\Gamma_{\mathcal{P}}(\frac{q}{m}\tau-4a)d\tau}\chi_{0}\big(4a-\frac{2aq}{k},-q\big)\nonumber\\
&+2\cos\left(\frac{aq}{\hbar}\right)\chi_{0}\big(-\frac{2aq}{k},-q\big)\Big]
\end{align}
while, in the energy decoherence process, it results in:
\begin{align}\label{probm}
P^{(\mathcal{E})}_\textrm{scr}(x)&=  \frac{2}{h^2}\int dq dp ds' \: e^{\frac{i}{\hbar}q x} e^{2 \Gamma_{\mathcal{E}} (p-\frac{q}{2},p+\frac{q}{2})t_\textrm{ref}}\cdot\nonumber\\
&\cdot e^{\frac{ip}{\hbar}(\frac{2aq}{k}-s')} \Big[\cos\Big(\frac{4ap}{\hbar}\Big)+\cos\Big(\frac{aq}{\hbar}\Big)\Big]\chi_0(s',-q)
\end{align}
Finally, Eqs.(\ref{probp}) and (\ref{probm}) can be used to estimate the visibility in Eq.~\eqref{eq:vis} given the explicit form of the state at the first beam splitter $\chi_0(s,q)$. We choose it to be a gaussian wavepacket of spread $\sigma$ in a superposition of momenta $\pm k$~\footnote{For the sake of clarity this corresponds to $\psi_0(x) =\sqrt{\frac{1}{2\sqrt{\pi}\sigma [1+\exp(\frac{-k^2\sigma^2}{\hbar^2})]}} \exp(-\frac{x^2}{2\sigma^2})\cos \Big(\frac{k x}{\hbar}\Big)$ in terms of the wavefunction.},
\begin{align}
\chi_{0}(s,q)=\frac{e^{-\frac{q^2 \sigma ^2}{4 \hbar^2}-\frac{s^2}{4 \sigma ^2}} \left(e^{-\frac{k^2 \sigma ^2}{\hbar^2}} \cosh \left(\frac{k q \sigma ^2}{\hbar^2}\right)+\cos \left(\frac{k
   s}{\hbar}\right)\right)}{e^{-\frac{k^2 \sigma ^2}{\hbar^2}}+1},
  \end{align}
With this initial state, the time at which the reflection occurs trivially reads $t_\textrm{ref} = a/v $ where $v=k/m$. The resulting formulas for the visibility are both very complicated \cite{supplemental}. However, in the case of positional decoherence, the formula can be simplified exploiting the fact that in the longitudinal-eikonal approximation the spread of the wavepacket is much smaller than the arm of the interferometer, i.e. $\sigma \ll a$. The visibility then simply reads:
\begin{equation}\label{all}
    \nu^{(\mathcal{P})} \simeq \exp\left(-\frac{4 a^3 \alpha ^2 c^3 m^3}{3 k L \hbar ^2} \right)
\end{equation}
This formula shows a reduction of the visibility proportional to square of the mass of the particle and to the cube of the interferometer's arm size, meaning that a small increase in the both of them will give an important gain in the sensitivity to spacetime fluctuations, if only decoherence in position occurs.\\
To make the study more concrete, we apply our results to a specific experiment, the HYPER interferometer, as it is a neat example of possible near future application of atomic Mach-Zehnder interferometers in space. Our goal is to study HYPER's sensitivity to scalar and tensorial metric fluctuations, i.e. those generated respectively by the $h_{00}$ and $h^{ij}$ terms of the metric. There is another study~\cite{lamine} in the literature concerning the effects of a stochastic gravitational perturbation on HYPER, which however deals specifically with the long wavelength tensorial perturbations constituting the so called Binary Confusion  Background. \\
We accordingly set the parameters of our simulated experiment as reported in Table~\ref{tab1}.
\begin{table}[h!]
\caption{\label{tab1}Parameters of the simulation.}
\begin{ruledtabular}
\begin{tabular}{l l l l l }
\textbf{m} [$\textrm{kg}$] &  \textbf{k} [$\frac{\textrm{kg}\, \textrm{m}}{\textrm{s}}]$ & \textbf{a} [$\textrm{m}$]  \\
$2.5*10^{-25} $ &  $8.8*10^{-28} $ &  $2.5*10^{-3}$  \\
\end{tabular}
\end{ruledtabular}
\end{table}\\ 
We analyse the sensitivity of the experiment for different values of the coherence length $L$ and the strength $\alpha$ of the fluctuations in the pure positional and energy decoherence cases. The main results are summarized by Fig.~(\ref{vis}, \ref{vism}) respectively.\\
\\
Our analysis shows that HYPER is sensitive to scalar fluctuations with correlation length and strength respectively down to $L \simeq 10^{-1} \textrm{m}$, and $\alpha \sim 10^{-20}$. These numbers can be improved to $L \simeq 10^{-3} \textrm{m}$, and $\alpha \sim 10^{-21}$ upon relaxing the assumption of long correlation length \cite{supplemental}.\\
In the case of the tensorial perturbation, we had to resort to a numerical analysis, as we were not able to find an analytic expression for the probability density and therefore the visibility \cite{supplemental}. Nevertheless, we have obtained the same numerical plot for the probability density for all possible choices for the strength $\alpha$ and correlation length $L$ of the perturbation in the range of interest: $10^{-40} < \alpha < 10^{-5}$, $1\: \textrm{m} < L < 10^{15}\: \textrm{m}$, as well for the decoherence free case ($\alpha = 0$). This means that the visibility is constant (within the numerical error) and $\nu^{(\mathcal{E})} = 1$. We conclude that HYPER is practically blind to tensorial perturbations in the long correlation length regime. This result is qualitatively in agreement with that obtained in \cite{lamine}. \\
The study so far was carried out assuming no other source of decoherence except gravitational fluctuations, while in real experiments different sources of decoherence are always present \cite{schlosshauer2007decoherence}.
We show that the most relevant source of decoherence, i.e. thermal gas collisions, gives a negligible effect. We will not consider other sources of decoherence because they strongly depend on the specific setup.
The decoherence function $\Gamma_{\textrm{coll}}(x-x)$ describing gas collision can be quite complex \cite{schlosshauer2007decoherence}, 
however in an interferometric experiment usually the superposition distance is much bigger than the typical thermal De Broglie wavelength of the gas allowing one to rely on the simplified expression: 
\begin{align}
\Gamma_{\textrm{coll}}= \frac{4\pi \Gamma(9/10)}{5 \sin(\pi/5)}\left(\frac{9\pi  \beta_{c} \beta_{g} I_{g} I}{64 \hbar \epsilon_{0}(I+I_{g})} \right) \frac{p_{g}v_{g}^{3/5}}{K_{b} T_{g}}
\end{align}
where $T_{g}$,$p_{g}$, $m_{g}$ are the temperature, the pressure and the mass of the gas, $I$, $I_{g}$ are the ionization energies, $\beta_c$ and $\beta_{g}$ the static polarizabilities of the matter-wave and gas particle and  $v_{g}= \sqrt{2 K_{b} T_{g}/m_{g}}$ is the thermal velocity of the gas particle.\\
Upon pluging in the values of the parameter relative to the experiment, which are summarized in Table \ref{tab2}, we get $\Gamma_{\textrm{coll}} \simeq 7.6 * 10^{-30} s^{-1}$, which shows that the decoherence induced by thermal gas collisions is indeed negligible with respect to gravitational decoherence on the sensitivity curve of Fig.(\ref{vis}).
\begin{table}[h!]
\caption{\label{tab2} Collisional parameters.}
\begin{ruledtabular}
\begin{tabular}{l l l l l }
$\mathbf{I_g}\;[ \textrm{eV}]$ & $\bold{\beta_g}\;[\textrm{m}^3]$ & $\mathbf{T_g}\;[\textrm{K}]$ & $\mathbf{p_g}\; [\textrm{Pa}]$\\
$13.6 $ &  $7.42*10^{-41}$ & $20$ & $10^{-11}$\\
\hline
$\mathbf{I_c}\; [\textrm{eV}]$ & $\mathbf{\beta_c}\;[ \textrm{m}^3]$\\
 $3.89$ &  $59.42*10^{-30}$\\
\end{tabular}
\end{ruledtabular}
\end{table}
\\
\textit{Summary.} - We have presented a general model describing the dynamics of non relativistic quantum matter subject to weak spacetime fluctuations.We have shown that the effect of such fluctuations can result in both position and/or momentum decoherence depending on the specific form of the fluctuations and the state of the quantum system, thus solving the preferred basis puzzle.  \\
We have then studied the effect of gravitational decoherence on a Mach-Zehnder interferometer, providing a practical formula to estimate the sensitivity of such a class of experiments to stochastic scalar fluctuations of the metric. We have also confirmed previous results showing how the HYPER proposed experiment is unable to detect stochastic tensorial metric fluctuations.\\
We have analysed the most relevant competing decoherence effect, namely thermal gas collisional decoherence, and shown that it is negligible with respect to gravitational decoherence.\\
Although based on strongly simplifying assumptions, this study shows that matter-wave interferometry is a promising avenue for testing interface of quantum mechanics and gravity and for the detection of the scalar gravitational background.
{\color{magenta} 
}

\section*{Acknowledgments}
LA and GG  deeply thank A. Belenchia, P. Creminelli, J.L. Gaona Reyes, A. Gundhi, C.I. Jones and K. Skenderis for the helpful and inspiring discussions. The authors acknowledge financial support form the EU Horizon 2020 research and innovation program under Grant Agreement No. 766900 [TEQ]. LA and AB thank the University of Trieste, INFN and the COST action 15220 QTSpace. GG thanks the Leverhulme Trust [RPG- 2016-046].
\bibliographystyle{unsrt}
\bibliography{biblio}

\begin{thebibliography}{10}

\bibitem{ligo}
The Ligo~collaboration the~virgo collaboration.
\newblock Observation of gravitational waves from a binary black hole merger.
\newblock {\em Physical Review Letters}, 116:061102, Feb 2016.

\bibitem{neutronstar}
LIGO~Scientific Collaboration and Virgo Collaboration.
\newblock Gw170817: Observation of gravitational waves from a binary neutron
  star inspiral.
\newblock {\em Physical Review Letters}, 119:161101, Oct 2017.

\bibitem{tian}
J.~Luo et~al.
\newblock {TianQin}: a space-borne gravitational wave detector.
\newblock {\em Classical and Quantum Gravity}, 33(3):035010, jan 2016.

\bibitem{decigo}
S.~Kawamura et~al.
\newblock The japanese space gravitational wave antenna - {DECIGO}.
\newblock {\em Journal of Physics: Conference Series}, 122:012006, jul 2008.

\bibitem{et}
M.~Punturo et~al.
\newblock The {Einstein} telescope: a third-generation gravitational wave
  observatory.
\newblock {\em Classical and Quantum Gravity}, 27(19):194002, sep 2010.

\bibitem{kagra}
The~KAGRA Collaboration.
\newblock {Construction of KAGRA: an underground gravitational-wave
  observatory}.
\newblock {\em Progress of Theoretical and Experimental Physics}, 2018(1), 01
  2018.

\bibitem{lisa}
K.~Danzmann and the LISA~study team.
\newblock {LISA}: laser interferometer space antenna for gravitational wave
  measurements.
\newblock {\em Classical and Quantum Gravity}, 13(11A):A247--A250, Nov 1996.

\bibitem{allen}
B.~Allen.
\newblock The stochastic gravity-wave background: Sources and detection.
\newblock {\em Relativistic Gravitation and Gravitational Radiation;
  Proceedings of the Les Houches School of Physics 26 Sept - 6 Oct, edited by
  J.A. March and J.P. Lasota, Cambridge University Press}, page 373, 1997.

\bibitem{qg1}
M.~Bojowald.
\newblock How quantum is the big bang?
\newblock {\em Physical Review Letters}, 100:221301, Jun 2008.

\bibitem{qg2}
A.~Ashtekar, T.~Pawlowski, and P.~Singh.
\newblock Quantum nature of the big bang.
\newblock {\em Physical Review Letters}, 96:141301, Apr 2006.

\bibitem{fuentes}
M.~Ahmadi C.~Sab{\'{\i}}n, D.E.~Bruschi and I.~Fuentes.
\newblock Phonon creation by gravitational waves.
\newblock {\em New Journal of Physics}, 16(8):085003, Aug 2014.

\bibitem{hogan}
J.~M. Hogan, D.~M.~S. Johnson, S.~Dickerson, T.~Kovachy, A.~Sugarbaker,
  S.~Chiow, P.~W. Graham, M.~A. Kasevich, B.~Saif, S.~Rajendran, P.~Bouyer,
  B.~D. Seery, L.~Feinberg, and R.~Keski-Kuha.
\newblock Observation of gravitationally induced quantum interference an atomic
  gravitational wave interferometric sensor in low earth orbit (agis-leo).
\newblock {\em General Relativity and Gravitation}, 43:1953--2009, Jul 2011.

\bibitem{gravwav}
E.Plagnol G.~Auger.
\newblock {\em An Overview of Gravitational Waves}.
\newblock World Scientific, 2017.

\bibitem{gao}
D.~Gao, P.~Ju, B.~Zhang, and M.~Zhan.
\newblock Gravitational-wave detection with matter-wave interferometers based
  on standing light waves.
\newblock {\em General Relativity and Gravitation}, 43:2027, Apr 2011.

\bibitem{cow}
R.~Colella, A.~W. Overhauser, and S.~A. Werner.
\newblock Observation of gravitationally induced quantum interference.
\newblock {\em Physical Review Letters}, 34:1472--1474, Jun 1975.

\bibitem{nonmin}
M.P. Hertzberg.
\newblock On inflation with non-minimal coupling.
\newblock {\em Journal of High Energy Physics}, 2010:23, Nov 2010.

\bibitem{astr}
B.~F. Schutz.
\newblock Gravitational wave astronomy.
\newblock {\em Classical and Quantum Gravity}, 16(12A):A131--A156, nov 1999.

\bibitem{qg}
C.~Kiefer.
\newblock {\em Quantum gravity}.
\newblock Oxford University Press, 2007.

\bibitem{hawk}
S.~W. Hawking.
\newblock The unpredictability of quantum gravity.
\newblock {\em Communications in Mathematical Physics}, 87(3):395--415, Dec
  1982.

\bibitem{diosi}
A.~Tilloy and L.~Di\'osi.
\newblock Principle of least decoherence for newtonian semiclassical gravity.
\newblock {\em Physical Review D}, 96:104045, Nov 2017.

\bibitem{classch}
N.~Altamirano, P.~Corona-Ugalde, R.~B. Mann, and M.~Zych.
\newblock Gravity is not a pairwise local classical channel.
\newblock {\em Classical and Quantum Gravity}, 35(14):145005, Jun 2018.

\bibitem{nimm}
K.~E. Khosla and S.~Nimmrichter.
\newblock Classical channel gravity in the newtonian limit.
\newblock {\em arXiv:1812.03118 [quant-ph]}, Dec 2018.

\bibitem{aha}
A.~Stern Y.~Imry, Y.~Aharonov.
\newblock Phase uncertainty and loss of interference: A general picture.
\newblock {\em Physical Review A}, 41:3436--3448, Apr 1990.

\bibitem{linet}
B.~Linet P.~Tourrenc.
\newblock Changement de phase dans un champ de gravitation- possibilité de
  détection interférentielle.
\newblock {\em Canadian Journal of Physics}, 54:1129, 1976.

\bibitem{sanchez}
J.L.~Sanchez Gomez.
\newblock Decoherence through stochastic fluctuations of the gravitational
  field.
\newblock {\em (ed.L. Diosi, B. Lukacs), pp. 88-93. Singapore: World
  Scientific}, 456:88--93, 1992.

\bibitem{power}
W.L.~Power I.C.~Percival.
\newblock Decoherence of quantum wave packets due to interaction with conformal
  space-time fluctuations.
\newblock {\em Proceedings: Mathematical, Physical and Engineering Sciences},
  456:955--968, 2000.

\bibitem{blencowe}
M.P. Blencowe.
\newblock Effective field theory approach to gravitationally induced
  decoherence.
\newblock {\em Physical Review Letters}, 111:021302, Jul 2013.

\bibitem{breuer}
H.P.~Breuer C.~L\"{a}mmerzahl, E.~G\"{o}kl\"{u}.
\newblock Metric fluctuations and decoherence.
\newblock {\em Classical and Quantum Gravity}, 26(10):105012, Apr 2009.

\bibitem{ana}
B.~L.~Hu C.~Anastopoulos.
\newblock A master equation for gravitational decoherence: probing the textures
  of spacetime.
\newblock {\em Classical and Quantum Gravity}, 30(16):165007, Jul 2013.

\bibitem{bosons}
L.~Asprea G. Gasbarri~A. Bassi.
\newblock Gravitational decoherence: a general non relativistic model.
\newblock {\em arXiv:1905.01121v2}.

\bibitem{foldy}
L.L.~Foldy S.~A.Wouthuysen.
\newblock On the dirac theory of spin 1/2 particles and its non-relativistic
  limit.
\newblock {\em Physical Review}, 78:29--36, Apr 1950.

\bibitem{weinberg}
S.~Weinberg.
\newblock {\em Cosmology}.
\newblock Oxford University Press, 2008.

\bibitem{hyper}
C.~Jentsch, T.~M{\"u}ller, E.~M. Rasel, and W.~Ertmer.
\newblock Hyper: A satellite mission in fundamental physics based on high
  precision atom interferometry.
\newblock {\em General Relativity and Gravitation}, 36(10), Oct 2004.

\bibitem{gallis}
G.N.~Fleming M.~R.~Gallis.
\newblock Environmental and spontaneous localization.
\newblock {\em Physical Review A}, 42:38--48, Jul 1990.

\bibitem{joos}
E.~Joos and H.D. Zeh.
\newblock The emergence of classical properties through interaction with the
  environment.
\newblock {\em Zeitschrift f{\"u}r Physik B Condensed Matter}, 59(2):223--243,
  Jun 1985.

\bibitem{talbot}
B.~Brezger, L.~Hackerm\"uller, S.~Uttenthaler, J.~Petschinka, M.~Arndt, and
  A.~Zeilinger.
\newblock Matter-wave interferometer for large molecules.
\newblock {\em Physical Review Letters}, 88:100404, Feb 2002.

\bibitem{wigner}
W.B. Case.
\newblock Wigner functions and {Weyl} transforms for pedestrians.
\newblock {\em American Journal of Physics}, 76(10):937--946, 2008.

\bibitem{wigrep}
V~I Tatarski{\u{\i}}.
\newblock The {Wigner} representation of quantum mechanics.
\newblock {\em Soviet Physics Uspekhi}, 26(4):311--327, apr 1983.

\bibitem{nimmrichter2014macroscopic}
S.~Nimmrichter.
\newblock {\em Macroscopic matter wave interferometry}.
\newblock Springer, 2014.

\bibitem{supplemental}
See supplemental material.

\bibitem{lamine}
B.~Lamine S.~Reynaud, M.T.~Jaekel.
\newblock Gravitational decoherence of atomic interferometers.
\newblock {\em The European Physical Journal D - Atomic, Molecular, Optical and
  Plasma Physics}, 20(2):165--176, Aug 2002.

\bibitem{schlosshauer2007decoherence}
M.~Schlosshauer.
\newblock {\em Decoherence and the Quantum-to-Classical Transition}.
\newblock Springer-Verlag Berlin Heidelberg, 2007.

\end{thebibliography}
\end{document}